\UseRawInputEncoding
\documentclass[aip,jcp,reprint,floatfix]{revtex4-1}
\usepackage{verbatim}%
\usepackage{amsmath,amsfonts} %
\usepackage[acronym]{glossaries} %
\usepackage{graphicx} %
\usepackage{dcolumn} %
\usepackage{bm} %
\usepackage{ifpdf} %
\usepackage{color} %
\usepackage{nicefrac} %
\usepackage{natbib} %
\usepackage[byname]{smartref} %
\usepackage[breaklinks, %
colorlinks, %
linkcolor=blue, %
citecolor=blue, %
urlcolor=blue]{hyperref} %
\usepackage[table]{xcolor} %
\usepackage{braket} %
\usepackage{parskip} %
\newcommand{\bea}{\begin{eqnarray}}
\newcommand{\eea}{\end{eqnarray}}

\def\be{\begin{equation}}
\def\ee{\end{equation}}
\def\bea{\begin{eqnarray}}
\def\eea{\end{eqnarray}}
\def\bal#1\eal{\begin{align*}#1\end{align*}}
\def\ba#1\ea{\begin{align}#1\end{align}}

\ifpdf %
 \fi

\begin{document}
                           
\sloppy
\title{
Utilizing relativistic time dilation for time-resolved studies
}

\author{Hazem Daoud}
\email {hazem.daoud@mail.utoronto.ca}
\affiliation{Department of Physics, University of Toronto, Toronto, Ontario, M5S 1A7, Canada}
\author{R. J. Dwayne Miller}
\affiliation{Department of Physics, University of Toronto, Toronto, Ontario, M5S 1A7, Canada}
\affiliation{Department of Chemistry, University of Toronto, Toronto, Ontario, M5S 3H6, Canada}

\date{\today}

\begin{abstract}
Time-resolved studies have so far relied on rapidly triggering a photo-induced dynamic in chemical or biological ions or molecules and subsequently probing them with a beam of fast moving photons or electrons that crosses the studied samples in a short period of time. Hence, the time resolution of the signal is mainly set by the pulse duration of the pump and probe pulses. In this paper we propose a different approach to this problem that has the potential to consistently achieve orders of magnitude higher time resolutions than what is possible with laser technology or electron beam compression methods. Our proposed approach relies on accelerating the sample to a high speed to achieve relativistic time dilation. Probing the time-dilated sample would open up previously inaccessible time resolution domains.

\end{abstract}
\maketitle
\glsresetall

\section{Introduction}
Until a few decades ago capturing molecular dynamics was in the realm of Gedanken- or thought experiments.\cite{Miller1,Miller2} In a chemical reaction chemists knew about the reactants and the final product but how the molecules and atoms rearranged themselves to produce the reaction products was always left to the realm of imagination. The reason for that is the difficulty in making these measurements. Chemical reactions typically happen at the speed of sound in solids ($\sim$1000 m/s)  and the atomic bond length is on the order of one angstrom and hence, the time resolution required is in the realm of femtoseconds and the spatial resolution is in the realm of angstroms.\cite{Sciaini,polanyi_direct_1995} \newline
The technological advancements of laser technology in recent decades have made it possible to produce laser pulses in the femtosecond and even in the attosecond regime.\cite{attosecond_science} This has enabled rapid developments in the field of ultrafast science. Typically, a short laser pump pulse is used to trigger some photo-induced reaction dynamic in a molecular sample that is subsequently probed by a probe pulse. The probe pulse is typically a short x-ray pulse\cite{Barends} in X-ray free electron laser (XFEL)\cite{altarelli_xfel_2006} facilities or a compressed electron pulse\cite{electron_propagation} in table top experiments.\cite{Sciaini,water_ediff} Additionally, laser pulses are used as probes in table-top spectroscopy experiments that temporally probe molecules and atoms but don't provide the spatial resolution of x-rays.\cite{Spectroscopy_Zewail_1988, Spectroscopy_Zewail_1990} XFELs, which are multi-billion dollar facilities, require high operating costs and involve truly remarkable feet of engineering.\cite{SLAC} Electron beams are generated in table-top experiments. They are typically accelerated via a DC electric field for a short distance to avoid rapid expansion due to the Coulomb forces between the electrons\cite{aluminum_melting} or they are first accelerated via a DC field and then compressed via an RF field.\cite{dc_ac_electrongun,rf_gun_eindhoven} There are also designs where acceleration and compression take place through the same RF field.\cite{Daoud_E_Gun} Another approach is to employ relativistic electron sources that greatly reduce pulse broadening effects and hence, achieve high brightness and time resolution.\cite{three_MeV_electron_source,five_MeV_electron_source,gold_MeV_electron_source} In all cases a short probe pulse is produced. The probe pulse captures the molecular dynamics and produces a diffraction pattern. By varying the time delay between the pump and probe pulses the molecular dynamics at different time points is captured and a 'molecular movie' can be produced. Hence, the time resolution is mainly limited by the technology to produce ever shorter laser and electron pulses,\cite{Zewail_res_limits} both to rapidly trigger a photo-induced dynamic and to subsequently image it. Other approaches take this standard approach a bit further by dissecting the probe pulse to get better time resolutions. In the case of electrons, streak cameras that spatially separate a long electron beam into smaller portions, and hence, higher time resolutions,\cite{Badali_DwayneMiller} have been developed.\cite{streak_camera} Another proposed method, known as optical gating,\cite{hassan_high-temporal-resolution_2017} uses ultrashort laser pulses to dissect the electron beam and achieve a higher time resolution than originally achievable by the length and speed of the beam. 

\section{Method}
In this paper we propose an alternative method to study molecular and atomic dynamics in time-resolved diffraction or spectroscopy studies with higher time resolutions without relying on advancements in laser or electron beam technology. The idea is to slow down the 'internal clock' of the sample (charged molecule or ion) instead of shortening the probe pulse. This can be achieved by accelerating the sample to relativistic speeds, which can be realized in particle accelerators such as cyclotrons and synchrotrons. A sample that is accelerated to speed $v_{s}$ undergoes a slowing down of its 'internal clock' by a factor of $\gamma$, where  \be\gamma = 1/\sqrt{1-v_{s}^2/c^2}\ee relative to the lab frame irrespective of its velocity direction. Hence, the time resolution becomes a function of the sample's energy rather than mainly being reliant on the pump and probe pulse durations. This can easily unlock new time resolutions that have never been achieved before. 

\section {Experimental and Theoretical Considerations}
As with any novel method there are a list of new barriers and challenges that are to be overcome in order to successfully implement it experimentally. In this section we propose an experimental setup and discuss experimental challenges and limitations and the involved physics in light of our proposed setup.
\subsection{Experimental considerations}
\subsubsection{Setup}
\begin{figure}[h]
	\centering
		\includegraphics[height = 7 cm]{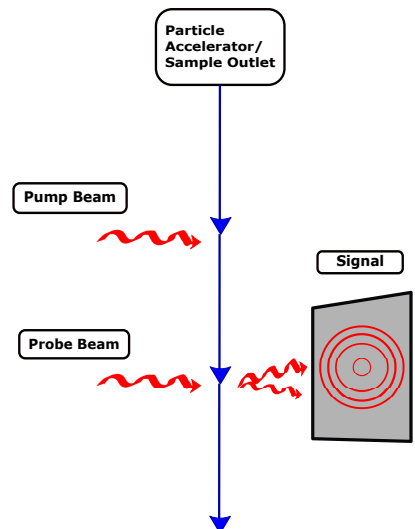}
	\caption{Schematic of the proposed experimental setup: Samples are accelerated to a fixed energy. A pump pulse and a probe pulse are directed parallel to each other and perpendicular to the sample direction of motion. The delay between the pump pulse and probe pulse can be controlled by changing the distance between the two beams.}
	\label{fig1}
\end{figure}

We propose accelerating the samples in a cyclotron or synchrotron and studying them at a fixed energy $E$, and hence, at a fixed speed $v_{s}$, that is not altered during data collection. Figure~\ref{fig1} shows a schematic of the experimental setup. Samples exit the acceleration phase into the chamber where the experiment is conducted. A pump pulse and a probe pulse are directed parallel to each other and perpendicular to the sample's direction of motion. The delay between the pump pulse and probe pulse can be controlled by changing the distance $L$ between the two beams. The resulting time delay $\tau^{'}_{d}$ according to the sample's clock is \be\tau^{'}_{d}=\frac{L}{v_{s}\gamma}.\ee The resultant signal will reflect the changing dynamics according to the 'internal clock' of the sample. 
\subsubsection{Sample suitability}
This novel method has its own set of unique challenges. To start with, it can only be applied to ions or molecules that are electrically charged so that they can be accelerated to relativistic speeds and are in the gas phase so they can reach the required energies. Moreover, the higher the mass of the sample the higher energy is required to run the experiment for a given time resolution. So, light, charged molecules and ions would be best suited for such studies. \newline
The typical number density in gas phase UED\cite{centurion_2016,yang_centurion_2015,UGED} and spectroscopy\cite{brünken_lipparini_stoffels_jusko_redlich_gauss_schlemmer_2019} experiments is $\sim$ $10^{15}$ cm$^{-3}$. Proton bunches at the LHC contain $1.15$ $\times$ $10^{11}$ protons with a proton number density of $\sim$ $10^{16}$ cm$^{-3}$.\cite{short_bunches,Steerenberg:2272573} Many bunch length reduction schemes have been proposed and indeed an order of magnitude shorter bunch length has been produced for the purposes of accelerating electrons using plasma wakefields of proton bunches.\cite{cern_wakefield}\newline
Another characteristic of a suitable sample is stability in the specific accelerator conditions. H- anions with a binding energy of $\sim$ 0.75 eV are accelerated regularly to 520 MeV at TRIUMF.\cite{TRIUMF} Covalent bonds of molecules typically have binding energies on the order of 1 eV or higher.\newline
Although originally designed to accelerate protons or positively charged ions only, the LHC ring has recently accelerated partially stripped Pb$^{+81}$ ions with one electron to an energy of 6.5 Q TeV,\cite{cern_lead} where Q is the ion charge number as part of the gamma factory proposal\cite{krasny2015gamma} to create a new type of high intensity light source. Engineering challenges with respect to collimation are currently being addressed.\cite{collimation_LHC}  
\subsection{Theoretical considerations}
\subsubsection{Energy considerations}
Currently, the Large Hadron Collider at CERN can accelerate protons to energies on the order of 7 TeV\cite{1811683} and lead ions to a collision energy of 5 TeV,\cite{CERN_lead_collision} which is enough to make remarkable gains in time resolution. As an example, a hydronium molecule ($H_{3}O^{+}$, rest mass: $3.16\times10^{-26}$ kg) accelerated to an energy of $1.8$ TeV would experience a slowing down of time with a $\gamma$ factor of $100$. Since \be E=\gamma mc^{2},\ee the time resolution scales proportionally to energy, so an energy of $18$ TeV would result in an astonishing $\gamma=1000$. Also, the time resolution is inversely proportional to mass so hydrogen ions, for example, would experience an order of magnitude higher gain in time resolution than hydronium molecules with the same energy.\newline
Even if these particle accelerators cannot be commissioned to run this type of experiment in the near future it will still be extremely interesting to observe the effect of relativistic time dilation on dynamical processes. With current laser technology it is possible to observe changes in differential detection, with and without a perturbation, as small as $10^{-4}$ to $10^{-8}$\cite{shot_noise1,shot_noise2} using standard modulation techniques and photon detectors.  There has also been major advances in laser based particle accelerators up to field gradients as high as $100$ GeV/m\cite{wakefields_review, GeV_electrons,joshi_corde_mori_2020,nature_wakefield} that will soon enable particle kinetic energies up to $10$ GeV range or higher.  This degree of relativistic energy would lead to only modest time dilation compared to what can be achieved at particle accelerators. Nevertheless, this time retardation would be directly measurable and would provide an important test case for further advances in laser based particle acceleration with the goal of directly controlling the time variable, asymptotically approaching 'stopping' time. This control of the time variable has the potential of opening new avenues beyond simple imaging to driving novel dynamics that would otherwise be too rapid to control.

\subsubsection{Pump and probe beam dynamics}
\begin{figure}[h]
	\centering
		\includegraphics[height = 6.5 cm]{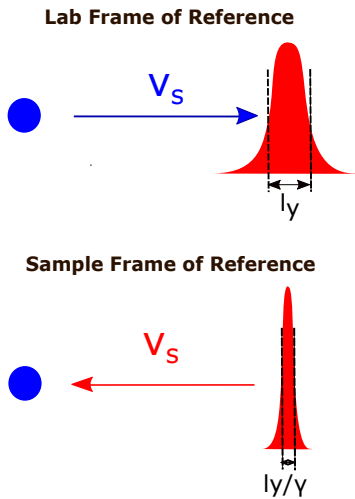}
	\caption{Lab and sample frames of reference: Along the $y$-direction the length of the pulse is contracted in the sample frame of reference relative to the lab frame of reference}
	\label{fig2}
\end{figure}

Pulsed pump and probe beams can be used as in standard ultrafast studies. The time resolution is mainly determined by the pulse duration of the pump and probe pulses. The pump pulse determines how fast a dynamic can be triggered and the probe pulse determines the imaging time resolution. The pulse duration conventionally refers to the time during which the full width at half max (FWHM) of the pulse crosses the sample. Other factors that affect time resolution are velocity mismatch\cite{charles_williamson_zewail_1993,goodno_astinov_miller_1999} that take place due to difference in velocity between the pump and probe pulses and due to their different incidence angles, and time of arrival jitter\cite{centurion_2016} for RF accelerated electron pulses. However, due to the experimental geometry lower resolution due to velocity mismatch or time of arrival jitter is avoided as both pulses are parallel so the delay time from time zero is solely determined by the speed of the sample between the pump and probe pulses. In our proposed setup the pulse crosses the sample in two directions, and hence, we will consider the pulse duration, during which the pulse crosses the sample or vice versa, in both directions. To explain the physics we denote the direction, parallel to the direction of propagation of the sample beam, $y$ and denote the perpendicular direction $x$. We will treat the problem from the lab frame of reference as well as from the sample frame of reference for clarification. \newline
\paragraph{Lab frame of reference}
The pulse duration in the $x$-direction $\tau_{x}$ is given by \be\tau_{x}=\frac{l_{x}}{v_{p}},\ee where $l_{x}$ indicates the pulse length in the $x$-direction and $v_{p}$ indicates the speed of the pump/probe pulse. The pulse duration in the $y$-direction $\tau_{y}$ is given by \be\tau_{y}=\frac{l_{y}}{v_{s}},\ee where $l_{y}$ indicates the pulse length in the $y$-direction and $v_{s}$ indicates the speed of the sample. In the proposed experiment $v_{s}$ will always be close to the speed of light $c$.  \newline 
\paragraph{Sample frame of reference}
As samples reach relativistic speeds close to the speed of light the effect of length contraction along the direction of sample propagation $y$ becomes significant. Hence, the length of the pulse in the $y$-direction is contracted to \be l^{'}_{y}=\frac{l_{y}}{\gamma}.\ee The pulse duration in the $y$-direction $\tau^{'}_{y}$ is then given by \be \tau^{'}_{y}=\frac{l_{y}}{\gamma v_{s}}.\ee The pulse duration in the $x$-direction $\tau^{'}_{x}$ is given by \be \tau^{'}_{x}=\frac{\gamma l_{x}}{v_{p}}\ee as $v_{p}$ transforms to \be v^{'}_{p}=\frac{v_{p}}{\gamma}\ee in the sample frame of reference. This is due to relativistic angle aberration. A pulse that is emitted at $90^{\circ}$ in the lab frame of reference does not hit the sample perpendicularly in the sample frame of reference. The closer $v_{s}$ is to $c$ the smaller is the incidence angle between the beam and the sample's line of motion in the sample frame of reference. Relativistic angular aberration will be discussed in more detail when discussing the observable signal. 
Typical pulses are Gaussian temporally and spatially ($\tau_{x}$ $\cong$ $\tau_{y}$) and so the time resolution $\tau_{res}$ would be determined by the relativistically shortened duration $\tau^{'}_{y}$. It would thus be given by \be\tau_{res}=\frac{1}{\gamma}\sqrt{\tau^{2}_{pump}+\tau^{2}_{probe}},\ee where $\tau_{pump}$ and $\tau_{probe}$ are the transit time durations of the sample beam through the pump and probe pulses in the lab frame, respectively. As an example for pump and probe beams with $l_{y}=10$ $\mu$m and $\gamma=100$, the time resolution would be roughly 470 as. Figure~\ref{fig2} shows a visual representation from both frames of reference along the relevant direction $y$.\newline


In conventional pump-probe experiments the lab frame of reference is the same as the sample frame of reference. A signal (e.g. diffraction pattern) always reflects the interaction time according to the clock rate of the sample and hence, our proposed setup takes advantage of the involved relativistic effects that result from the differences between the two frames of reference.

\subsubsection{Doppler effect and frequency shifts}
It is of significant importance to understand how the frequency of a laser, x-ray or electron pulse, which is created in the lab frame, is 'seen' by the sample in its own rest frame in order to properly carry out the experiment. For lasers, a significant shift in frequency may push it outside the absorption spectrum of the sample and hence, the intended interaction would not take place. For scattering x-rays and electrons, if they undergo significant redshift, for example, their spatial resolution would decrease. Relativistically, in addition to the classical Doppler effect, there is the added effect of time dilation. Even if the source and receiver are not crossing paths, relativity dictates a frequency shift known as transverse Doppler effect.\cite{Doppler_verification}\newline 
If we let $\theta$ be the angle between the sample wave vector and the wave vector of the pump/probe particles, as measured in the lab frame of reference, then the frequency that the sample 'sees', $f_{s}$, is given in terms of the frequency in the lab frame of reference, $f_{l}$, by \be f_{s} = f_{l}\gamma (1-\beta \cos{\theta})\ee for photons.\cite{Doppler_Li} For other particles, e.g. electrons, one needs to replace $\beta$ with \be\beta_{e}=\frac{v_{s}}{v_{e}},\ee where $v_{e}$ is the speed of the particles but $\gamma = 1/\sqrt{1-v_{s}^2/c^2}$ remains the same.
In our proposed setup, where the pump and probe beams are perpendicular to the direction of motion of the samples, we have  \be f_{s} = f_{l}\gamma.\ee  In fact, depending on the angle, there could either be a redshift or a blueshift. There is no frequency shift for one critical angle $\theta_{c}$. For $\beta$ ($\beta_{e}$) $\approx$ $1$, as $\gamma$ becomes larger, $\theta_{c}$ becomes smaller, meaning that the two wave vectors are more collinear. Although this would eliminate frequency shifts entirely, it would decrease the time resolution significantly.

\subsubsection{The observable signal}
Light can interact with matter in many different ways (e.g. absorption, scattering, etc.). In this section we present a general scheme for calculating the final observable signal in our proposed experiment.\newline
The main steps are: (1) Transforming the incident field from the lab frame of reference to the sample frame of reference by applying the Lorentz transformations; (2) Calculating the resultant signal in the sample frame of reference; (3) Transforming the signal from the sample frame of reference to the lab frame of reference by applying the Lorentz transformations one more time.\newline
Without loss of generality, for an incident electric field $E_{i}$ with angular frequency $\omega_{i}$ polarized in the z-direction and incident at angle $\theta_{i}$ (angle between photon wave vector $\vec{k_{i}}$ and sample velocity vector $\vec{v_{s}}$), the field would be Lorentz transformed to the sample frame of reference to $E^{'}_{i}$ in the following way:\cite{lee_mittra_1967, papas_1965}
\be\vec{E_{i}} = \hat{z} E_{i} \exp{[ik_{i} (\cos{\theta_{i} y}+sin{\theta_{i} x})]} \exp{(-i\omega_{i} t)},\ee 
\be\vec{E^{'}_{i}} = \hat{z}^{'} E^{'}_{i} \exp{[ik_{i}^{'} (\cos{\theta_{i}^{'} y^{'}}+sin{\theta_{i}^{'} x^{'}})]} \exp{(-i\omega^{'}_{i} t^{'})},\ee
where
\be E^{'}_{i}=\gamma(1-\beta \cos{\theta_{i}}) E_{i},\ee
\be k^{'}_{i}=\frac{\omega^{'}_{i}}{c},\ee
\be\omega^{'}_{i}=\gamma(1-\beta \cos{\theta_{i}}) \omega_{i},\ee
\be\theta^{'}_{i}=\cos^{-1}{[\frac{\cos{\theta_{i}}-\beta}{1-\cos{\theta_{i}}(\beta)}]}.\ee
For particles other than photons moving with speed $v_{e}$ the angular frequency and angle transform in the following way:
\be\omega^{'}_{i}=\gamma(1-\beta_{e} \cos{\theta_{i}}) \omega_{i},\ee
\be\theta^{'}_{i}=\tan^{-1}{[\frac{\sin{\theta_{i}}}{\gamma(\cos{\theta_{i}}-\beta_{e})}]}.\ee
However, as before $\gamma = 1/\sqrt{1-\beta^2}$ with $\beta=\frac{v_{s}}{c}$ remains the same. \newline
Due to the nature of the angular transformations, we expect scattering and diffraction angles to be wider than for the static case, and hence, we recommend detectors that cover as much of the 4$\pi$ sr solid angle as possible.

\section {Conclusion}
We proposed a new method for time resolved studies that relies on taking advantage of relativistic effects rather than on advances in laser and electron beam compression technology. We have shown that our method has the potential of improving time resolutions by $2$ or $3$ orders of magnitude using currently available technology. This has the potential of unlocking a whole new domain of ultrafast dynamics that was previously unattainable. In a future that is pointing towards ever more powerful accelerators there will be a huge potential to achieve truly remarkable time resolutions that exceed the status quo by orders of magnitude.  We hope this paper will open up new avenues of collaboration between the particle physics community and the ultrafast science community that would result in maximizing the research potential of particle accelerator facilities worldwide.

\section {Acknowledgments} 
H.D. would like to thank Prof. Pierre Savaria and Prof. David Bailey for fruitful discussions. 

\section {Data Availability} 
Data sharing is not applicable to this article as no new data were created or analyzed in this study.

\bibliography{mybib}

\newpage

\end{document}